\documentclass[%
reprint,
groupedaddress,
prstab,
]{revtex4-1}
\usepackage{amsmath}
\usepackage{amssymb}
\usepackage{graphicx}
\usepackage{dcolumn}
\usepackage{bm}

\begin{document}

\newcommand{\I}{\mathrm{i}}
\newcommand{\E}{\mathrm{e}}
\newcommand{\D}{\,\mathrm{d}}
\newcommand{\s}{\mathrm{sign}}
\newcommand{\sinc}{\mathrm{sinc}}

\title{Theoretical investigations on the Adiabatic Matching Device--based positron capture system}

\author{Eugene Bulyak}
\email{Corresponding author: Eugene.Bulyak@gmail.com; bulyak@ijclab.in2p3.fr; bulyak@kipt.kharkov.ua}
\altaffiliation[\\Also at: ]{Karazin National University, 4 Svobody sq., Kharkiv, Ukraine and Universit\'{e} Paris-Saclay, CNRS/IN2P3, IJCLab, Orsay, 91405, France}
\affiliation{NSC KIPT, 1 Academichna str, Kharkiv, Ukraine}

\author{Viktor Mytrochenko}
\altaffiliation[Also at: ]{Universit\'{e} Paris-Saclay, CNRS/IN2P3, IJCLab, Orsay, 91405, France}
\affiliation{NSC KIPT, 1 Academichna str, Kharkiv, Ukraine}

\author{ Iryna Chaikovska}
\email{Corresponding author: Iryna.Chaikovska@ijclab.in2p3.fr}
\affiliation{Universit\'{e} Paris-Saclay, CNRS/IN2P3, IJCLab, Orsay, 91405, France}

\author{Viacheslav Kubytskyi}
\affiliation{Universit\'{e} Paris-Saclay, CNRS/IN2P3, IJCLab, Orsay, 91405, France}

\author{Robert Chehab}
\affiliation{Universit\'{e} Paris-Saclay, CNRS/IN2P3, IJCLab, Orsay, 91405, France}

\author{Fahad Alharthi}
\affiliation{Universit\'{e} Paris-Saclay, CNRS/IN2P3, IJCLab, Orsay, 91405, France}

\begin{abstract}
The positrons produced with the electron beam impinging on a conversion target, possess wide energy spectrum and large sweep of the angle of trajectories to the system axis. Accommodation of the positron bunch to the acceptance of an ajacent accelerator, mandates the reduction of angular spread. One of the most appropriate devices for transforming the phase portrait of a positron bunch is Adiabatic Matching Device (AMD). The paper presents an abridge theory of AMD. It is shown that the transformation of the transverse phase phase volume aimed at decrease the angular spread causes prolonging the bunch. Both the longitudinal and the transversal probability density functions are derived. The analytical results are validated with numerical simulations.
\end{abstract}

\maketitle

\section{Introduction}
The Future Electron--Positron Circular Collider (FCCee) requires intense po\-si\-tron beam, \cite{bulyak19b,chaikovska19}.
The positrons produced with the electron beam possess wide energy spectrum and large sweep of the angle of trajectories to the system axis, at the same time a small transverse and longitudinal sizes. Accommodation of this bunch to the acceptance of an ajacent accelerator, mandates the reduction of angular spread.

One of the widely employed  devices for the beam transverse portrait transformation is Adiabatic Matching Device (AMD),  \cite{chehab92} and references therein.
This kind of the matching device also is intended to employ in the positron sources of other machines and projects, see \cite{wang08,zang12,zang14,bulyak20}.

\section{Adiabatic Matching Device}
\begin{figure}
\includegraphics[width=8.6cm]{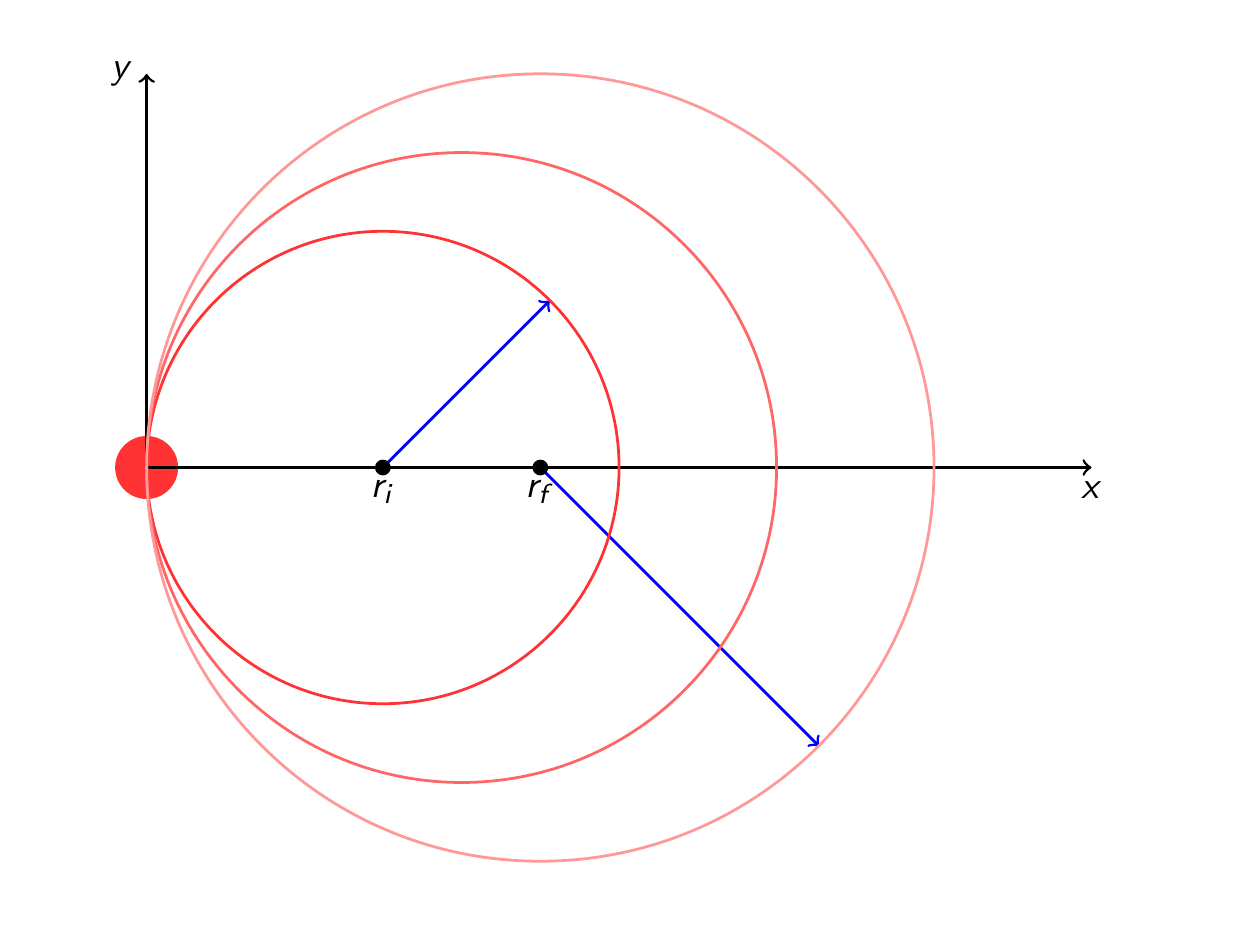}
\caption{Initial, intermediate, and final projections (Larmor rings).  \label{fig:traject}}
\end{figure}

\begin{figure}
\centering
\includegraphics[width=8.6cm]{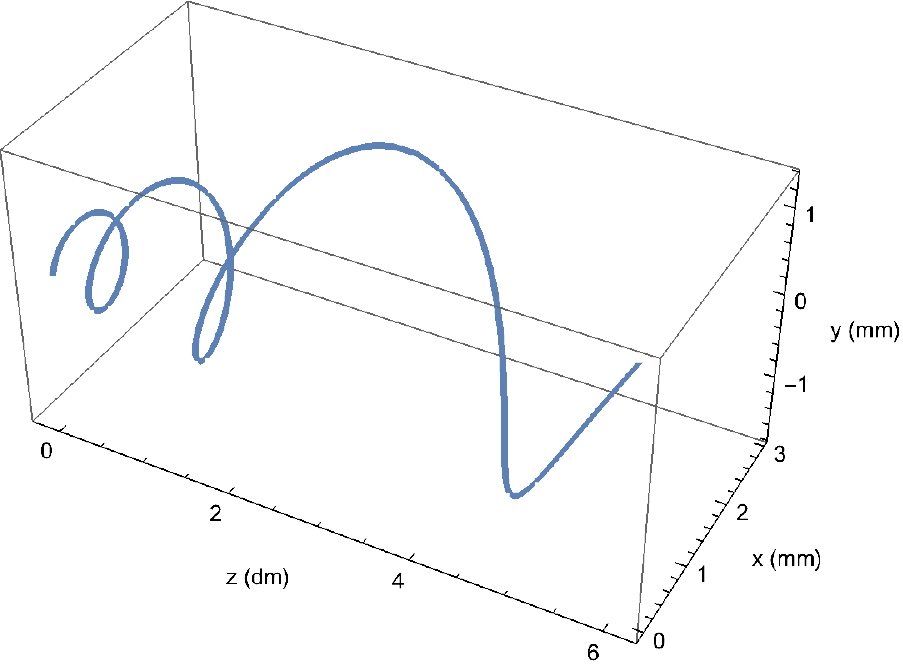} \\
\includegraphics[width=8.6cm]{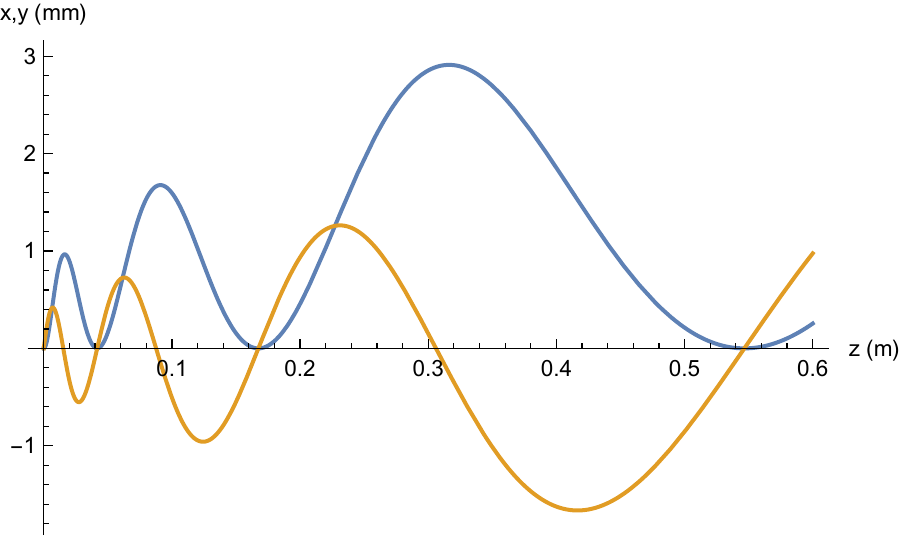}

\caption{Trajectory (top) and its projections upon $(x,0,z)$ -- blue -- and $(y,0,z)$ -- yellow -- planes; short AMD. \label{fig:xytraj}}
\end{figure}

The solenoidal magnetic field, which magnitude gradually decreased along the system axis $z$, is employed in AMD. Snapshots of projections of a positron trajectory on the transverse plane are presented on Fig.\ref{fig:traject} and projections of trajectories on the longitudinal planes on Fig.\ref{fig:xytraj}.

As it may be seen from these figures, the trajectories of positrons (and electrons as well) are more strighten at the exit of AMD while the amplitudes increased.    

\subsection{Brief  Theory of AMD}
Principle of the AMD operation is described in \cite{bulyak18a,bulyak20a,chaikovska22}.
Transformation of a positron bunch is processed while the positrons emitted from the conversion target are passing  through the tapered solenoidal magnetic field.
 
 The magnitude of the field gradually decreases along the beam's passage.
Magnetic field on the AMD axis $z$ is usually  taken in a form, see \cite{chehab92}.
\begin{equation} \label{eq:fieldz}
B_z(\rho = 0) = \frac{B_0}{1+\alpha z}\;,
\end{equation}
where $\alpha $ is  the tapering parameter, $z$ is the axial coordinate, $B_0$ is the field magnitude at the front end of the AMD, $z=0$.
This field in the cylindrical frame can be associated with a vector potential $A$ having only a nonzero $\theta $ component, \cite{bulyak20}:
\[
A=A_\theta = B_0\frac{\rho }{2(1+\alpha z)}\; ,
\] 
from which the radial component of the field yields 
\[
B_\rho(\rho ,z) = \frac{B_0 \alpha \rho }{2 (1+\alpha z)^2}\; .
\]
Here $\rho $ is the radial coordinate.

The positron trajectory in such a field is helical (spring-like), which  presents a uniform motion of the Larmor rings along the system axis. In the tapered field \eqref{eq:fieldz}, the Larmor radius  increases with the  distance traveled. 

Two parameters are used  as the basics for studying AMD. They are:
(i) the cyclotron frequency of  positrons in the field magnitude $B$, 
\begin{align} \label{eq:omegac}
\omega_\text{c} &= \frac{e B}{\gamma m} = \frac{e B}{\gamma m}\; ,
\intertext{and (ii) the radius of gyration (the radius of the Larmor ring)}
r & =  \frac{\gamma m c \sin \theta_0}{e B}\; , \label{eq:radlarm}
\end{align}
where  $c$ is the speed of light, $e$ is the positron charge, $m$ is the rest mass, $\gamma $ is the Lorentz factor, $\theta_0 $ is the initial angle between the positron trajectory and $z$ axis.

\subsection{Adiabatic Approach}
Since the positron oscillates many times on its way from the front end  to the exit of the AMD,  the adiabatic approach can be implemented, see  \cite{bakay81}. 
(The Bursch theorem, see, e.g., \cite{lawson77}, is a particular case of the general adiabatic theory.)
 Thus the cyclotron frequency \eqref{eq:omegac} is taken as decreasing with the field magnitude:
\[
\omega = \omega_0 / (1+\alpha z)\; ,
\] 
where $\omega_0 $ is the frequency in the field magnitude $B_0 $.

The adiabatic theory, \cite{bakay81}, provides that the adiabatic integral invariants remain constant when the system parameters vary slowly. We assume the rotational frequency of the positron in a slowly decreasing magnetic field varies slowly, for many periods (the period, in turn, varied as well).

Within paraxial approximation -- the transverse momentum is sufficiently less than the longitudinal one -- the `action' variable is an adiabatic invariant:
\[
J = \int_0^\tau E_\perp \D t = \int_0^\tau \frac{1}{2}\gamma m r^2 \omega^2 \D t = \text{const}\; ,
\]  
where $\tau $ is the period of rotation.

Since $\tau\propto 1/\omega $, we are concluding that
\begin{equation}\label{eq:r}
r^2 \omega = r^2 \frac{\omega_0}{(1+\alpha z )} = \text{const}  \Rightarrow r(z) = r_0 \sqrt{1+\alpha z }\; .
\end{equation}

Since the period is increasing as $1/\omega \propto (1+\alpha z)$, the angle between the trajectory direction and $z$-axis is decreasing as
\begin{equation}\label{eq:theta}
\theta (z) = \theta_0 /\sqrt{1+\alpha z}
\end{equation}

Expressions \eqref{eq:r} and \eqref{eq:theta} prove the main goal of the AMD: decreasing of the angular spread at a mutual increase of the bunch radius (adiabatic focusing).

\subsection{Trajectories Lengthening}
Adiabatic focusing has a side effect -- trajectory lengthening -- which causes the positron bunch to smear longitudinally. 

For arbitrary emitting angles, $0\le \theta_0 < \pi / 2$, and under the assumption of the AMD straightening of the angle,  \eqref{eq:theta},
the first fundamental form reads
\begin{align*}
\sqrt{1+\left(\frac{\D y}{\D z}\right)^2} &= \sqrt{1+\tan^2\left(\frac{\theta_0}{1+\alpha z}\right)} \\ &= \sec \left(\frac{\theta_0}{\sqrt{1+\alpha z}}\right)\; .
\end{align*}

Therefore, the trajectory started at the angle $\theta_0$,  by the rear end  of the AMD is longer as compared to the reference,  $\theta_0 = 0$, by
\begin{equation}\label{eq:arbbunchl}
\zeta (\theta_0; L)\equiv \Delta L = \int_0^L\sec \left(\frac{\theta_0}{\sqrt{1+\alpha z}}\right)\D z - L\;  ,
\end{equation}
where $L$ is the AMD length.

\begin{figure}
\includegraphics[width=8.6cm]{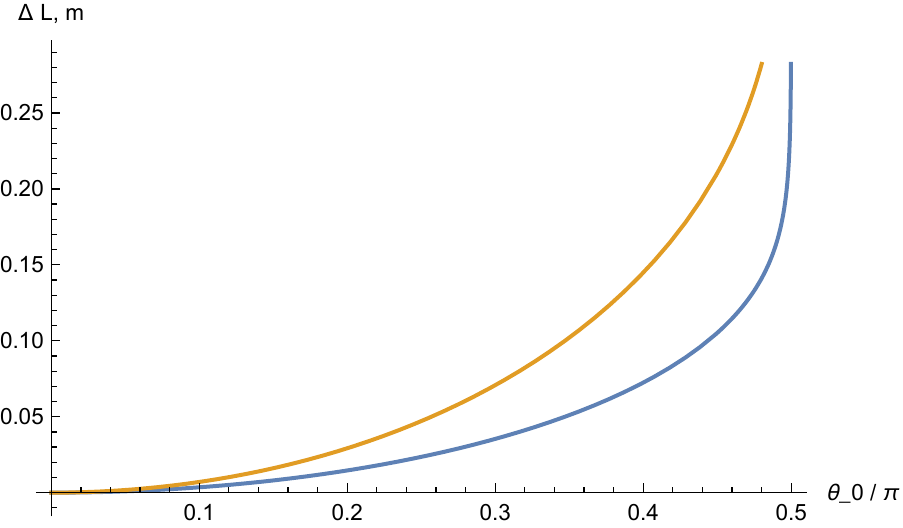}
\caption{Trajectory lengthening vs. initial angle for $L= 60\,\text{cm}$ (blue) and  $L= 120\,\text{cm}$  (yellow).   \label{fig:dellt}}
\end{figure}

A plot of the trajectory lengthening by the AMD $L = 60\,\text{cm}, 120\text{cm}$, $\alpha = 48.3, 24.15$ (reduction of the field magnitude from 15\,T to 0.5\,T for both cases) is presented in Fig.\ref{fig:dellt}.

As follows from the figure, a shorter AMD produces a shorter bunch length. On the other hand, the applicability conditions of the adiabatic approach become worse. 

\subsection{Probability Density Functions}
Evaluation of the longitudinal and the transversal PDFs for the bunch at AMD rear end will be performed under the assumption of the Rayleigh distribution for the initial angles $\theta_0 $  with the dispersion $\sigma $ as a natural:
\begin{equation}\label{eq:ray}
g(\theta_0; \sigma ) = \frac{\theta_0}{\sigma ^2}\, \exp \left( -\frac {\theta_0^2}{2 \sigma ^2} \right) \;  .
\end{equation}

\subsubsection{Longitudinal Bunch Profile}
In real devices, the angular spread reasonably accepted for further processing is relatively small, $\theta_0\ll 1$. Therefore, the integrand in \eqref{eq:arbbunchl} can be expanded in series around $0$:
\[
\sec \left(\frac{\theta_0}{\sqrt{1+\alpha z}}\right)\approx 1+\frac{\theta_0^2}{2(1+\alpha z)}\; ,
\]  
and \eqref{eq:arbbunchl} is reduced to
\begin{equation}\label{eq:arbbunchls}
\zeta = \int_0^L \frac{\theta_0^2}{2(1+\alpha z)}\D z = \frac{\theta_0^2}{2}\,\frac{\log (1+\alpha L)}{\alpha } \;  .
\end{equation}

For this case taking into account the Rayleigh distribution  of $\theta_0 $, see \eqref {eq:ray}, the PDF for $\zeta $ casts into
\begin{align}\label{eq:pdfzeta}
f(\zeta;\alpha, L) &= \frac{\alpha }{\sigma^2 \log (1+\alpha L)}\nonumber \\
&\times \exp\left[ - \frac{\zeta\alpha }{\sigma^2 \log (1+\alpha L)} \right]\; .
\end{align}

A positron bunch having $\delta $ distribution at the AMD front end, acquires an exponential longitudinal density distribution with the rate parameter $\lambda  $ at the rear end:
\[
f(\zeta;\lambda)  =\lambda \E^{-\lambda \zeta}\,;\quad \lambda = \frac{\alpha }{\sigma^2 \log (1+\alpha L)}\; .
\]     

Convolving \eqref{eq:pdfzeta} with an initial longitudinal Gaussian bunch profile, we get 
\begin{align}\label{eq:longau}
g_\text{long} (\zeta; s,\lambda) &= \frac{1}{2} \lambda  \exp \left[\frac{1}{2} \lambda  \left(\lambda  s^2-2 \zeta \right)\right] \nonumber \\
&\times \left[\mathrm{F}\left(\frac{\zeta
   -\lambda  s^2}{\sqrt{2} s}\right)+1\right] \; ,
\end{align}
where $s$ is the Gaussian dispersion in the initial bunch shape, and $\mathrm{F}(\cdot)$ is the error function.

\begin{figure} [htb]
\includegraphics[width=8.6cm]{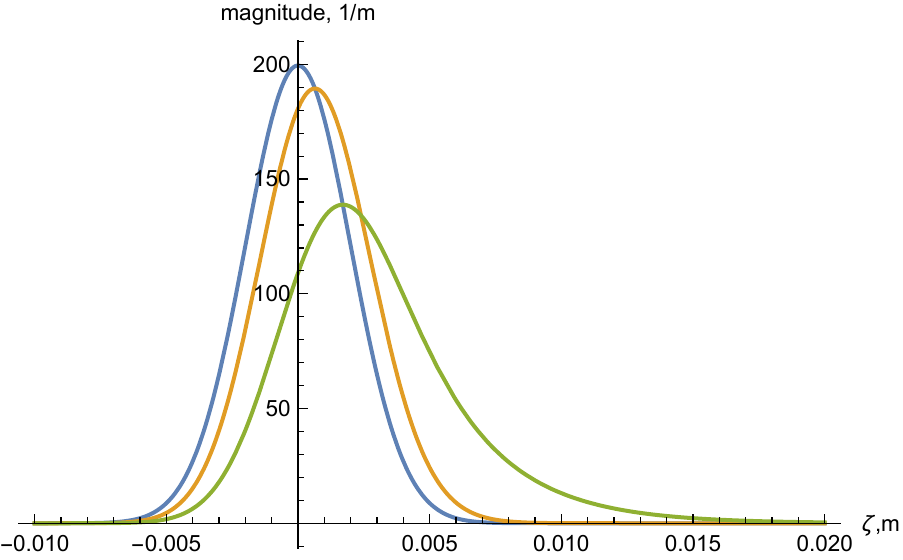}
\caption{Longitudinal probability density functions: initial PDF (blue), final PDFs:  $\sigma = 0.1$ (yellow) and $\sigma = 0.2 $ (green). $s = 0.002$\,m, $\alpha = 48.3\,\text{m}^{-1}$, $L = 0.6$\,m.   \label{fig:longpdf}}
\end{figure}

Being Gaussian at the AMD front end, the bunch dilutes  later to: the mode (maximum)  offset, and the backward tail becomes thicker. In figure \ref{fig:longpdf} the final PDFs for the two Rayleigh dispersions are presented.   

\subsubsection{Transverse Bunch Profile}
Since the radius of the Larmor ring is directly proportional to the initial angle $\theta_0$, 
\[
r_0=\theta_0 \frac{\gamma m c}{eB_0}\;; \Rightarrow \sigma_r = \sigma \frac{\gamma m c}{eB_0}\; ,
\]
the initial radius also possesses the Rayleigh distribution with the dispersion $\sigma_r $.

The radial distribution density of a ring with the radius $a$ offset from the axis by the same magnitude $a$ (see Fig.\ref{fig:traject}) reads:
\begin{equation}\label{eq:offring}
f(r,a)=\frac{\Theta (2 a-r,r)}{\pi  \sqrt{2 a r-r^2}}
\end{equation}
with $\Theta $ being the Heaviside step function.

We can convolute \eqref{eq:offring} with the Rayleigh distribution, then get:
\begin{widetext}
\begin{align}\label{eq:bess}
\mathcal{F}(r,\sigma_r ) &= \frac{1}{16 r \sigma_r ^2}\,\exp \left( - \nu \right)   \left\{r^2\left[ I_{-\frac{1}{4}}\left(\nu\right)-
   I_{\frac{1}{4}}\left(\nu\right)- I_{\frac{3}{4}}\left(\nu\right) + I_{\frac{5}{4}}\left(\nu\right)\right]
   + 8 \sigma_r ^2 I_{\frac{1}{4}}\left(\nu\right)\right\}\;  ;\\
   &\nu \equiv \frac{r^2}{16 \sigma_r ^2}\; , \nonumber
\end{align}
where $ I_n(\cdot ) $ is the modified Bessel function of the first kind.
\end{widetext}

This PDF may be converted to the Cartesian frame. In this frame, the RMS for one of the coordinates, say $x$, yields:
\[
\left< x^2\right> = \frac{3}{2}\,\sigma_r^2\; .
\] 

\begin{figure}
\includegraphics[width=8.6cm]{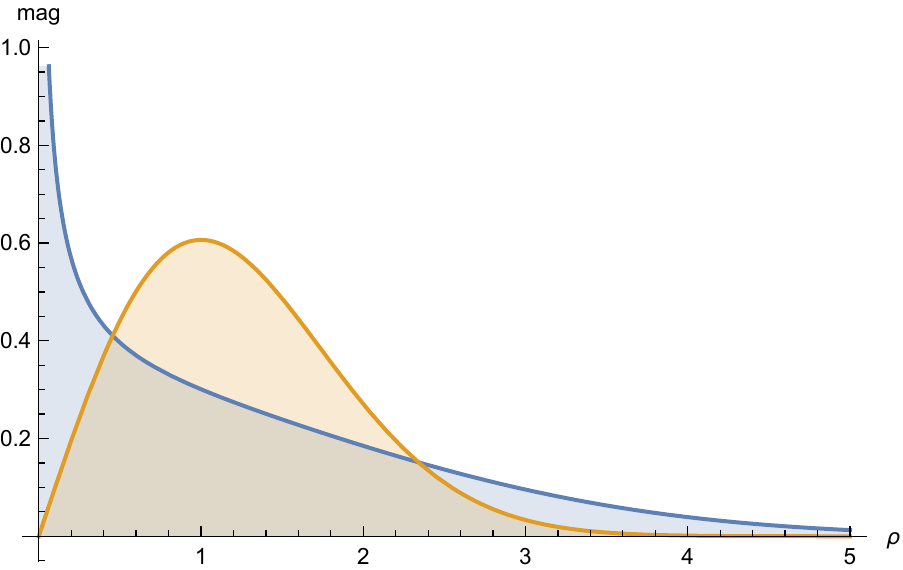}
\caption{Probability density distributions of Rayleigh (yellow) and \eqref{eq:bess} (blue).   \label{fig:raymy}}
\end{figure}

Despite of the approximately same RMS (factor of $\sqrt{3/2}$) PDFs differ one from another    significantly, see Fig.\ref{fig:raymy}.

\subsection{Simulations on AMD}
In order to validate the theory, simulations of the positron trajectories in AMD were done. The code ASTRA was used, \cite{floettmann03,floettmann17} .

Two cases were simulated: the `short' AMD, $L=0.6\,\text{m},\;\alpha = 48.3$, and the `long' AMD, $L=1.2\,\text{m},\;\alpha = 24.15$, with the same initial, $B_0 = 15\,\text{T}$, and final, $B_\text{fin} = B_0 / (1+\alpha L) =0.5\,\text{T}$, field magnitude.  

\subsubsection{Longitudinal bunch transformation}
Results of the theoretical evaluations together with the corresponding simulations are presented in the Table~\ref{tab:long}.

\begin{table}
\caption{Bunch RMS lengthening for
$\gamma = 32$, $B_0 = 15$\,T, $\sigma_\text{initial} = 0.88$\,mm \label{tab:long}}

\centering
\begin{tabular}{ccccc}
\hline
 $\sigma_\theta $ & $L$, m & $\alpha,\text{m}^{-1}$ & $\sigma_z^{(theo)} $, mm & $\sigma_z^{(sim)} $ , mm \\ 
 \hline
0.1 & 0.6 & 48.30 & 1.13 & 1.06 \\
0.1 & 1.2 & 24.15 & 1.66 & 1.50 \\
0.2 & 0.6 & 48.30 & 2.95 & 2.71 \\
0.2 & 1.2 & 24.15& 5.70 & 5.21 \\
\hline
\end{tabular}
\end{table}

Comparison of the theoretical predictions with the simulation results displays a good agreement between them. Bigger theoretical RMSs may be explained by more accurate accounts for the (infinite) tails in the theory, as the simulation involves relatively small number of particles, $N_\text{part } = 500$. 

\subsubsection{Transversal bunch transformation}
\begin{figure}
\includegraphics[width=8.6cm]{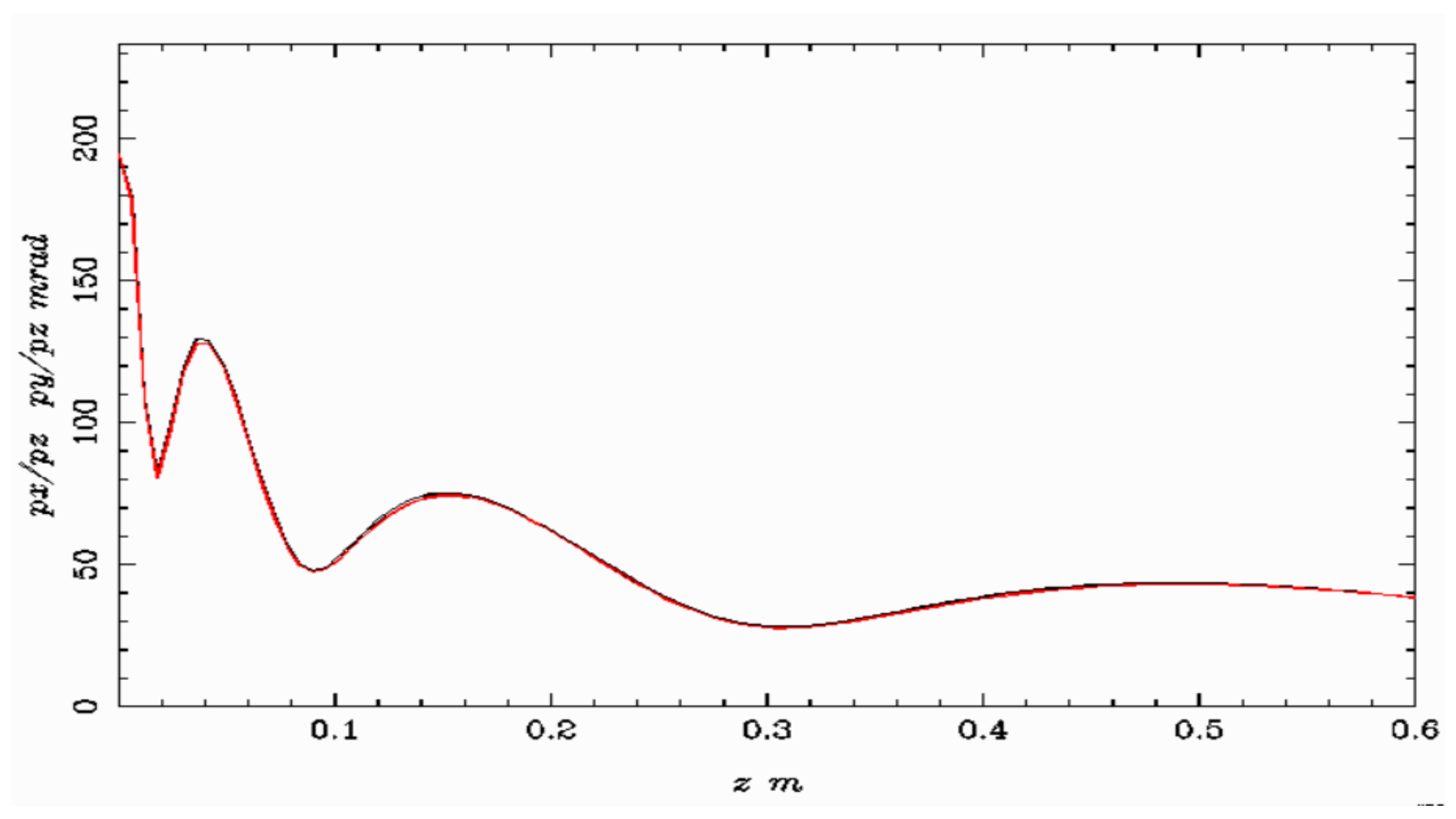}\\
\includegraphics[width=8.6cm]{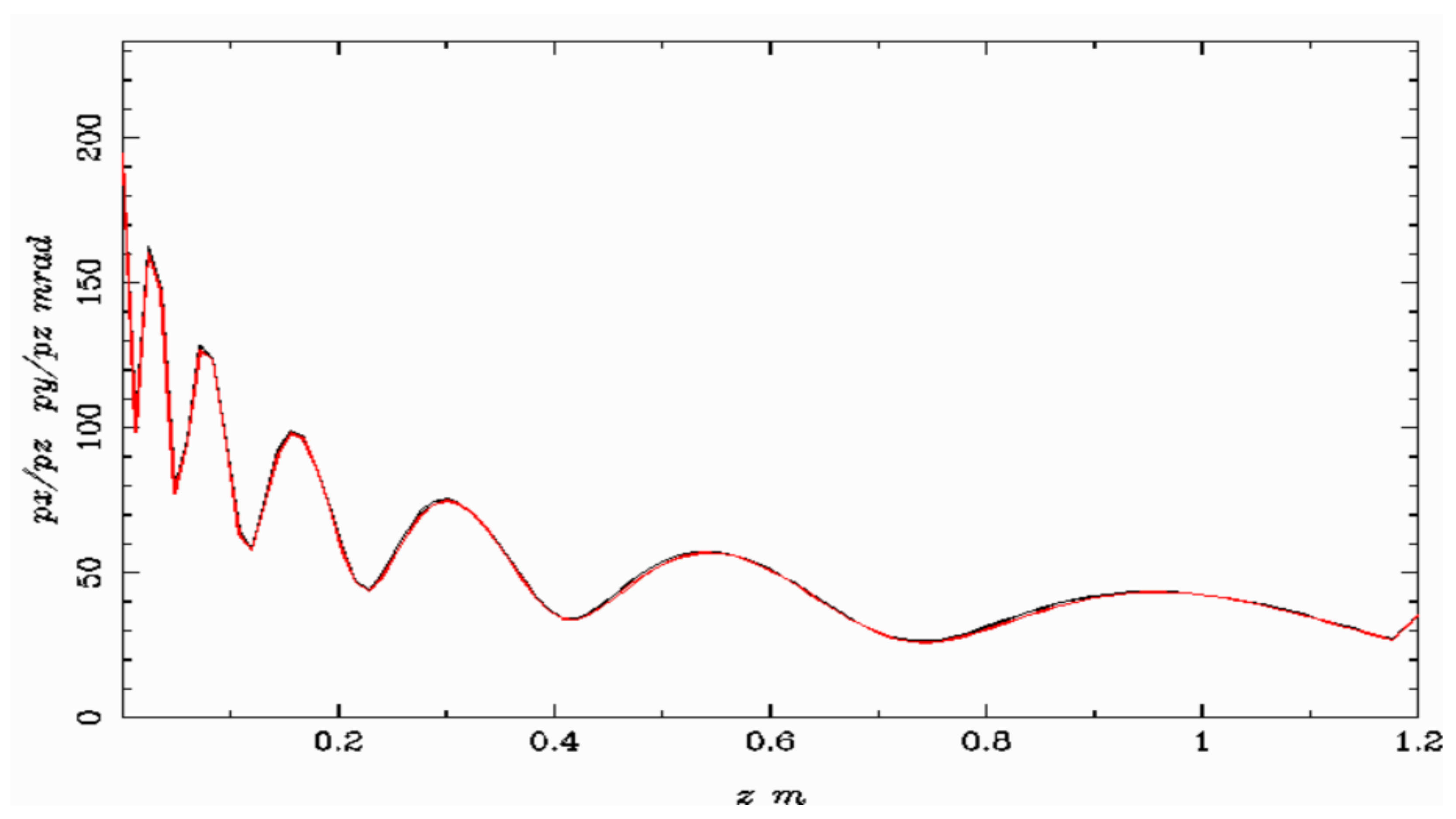}
\caption{The angular spread envelope evolution, $\alpha = 48.3\,\text{m}^{-1}$ (top) and $\alpha = 24.15\,\text{m}^{-1}$ (bottom). \label{fig:sigmaxp48}}
\end{figure}

Simulated angular RMS spread evolutions along $z$-axis are presented in Fig.~\ref{fig:sigmaxp48},  $B_0 = 15\,\text{T}$, $\gamma = 32 $ for different $\alpha $ (short and long AMD).

\begin{figure} 
\includegraphics[width=8.6cm]{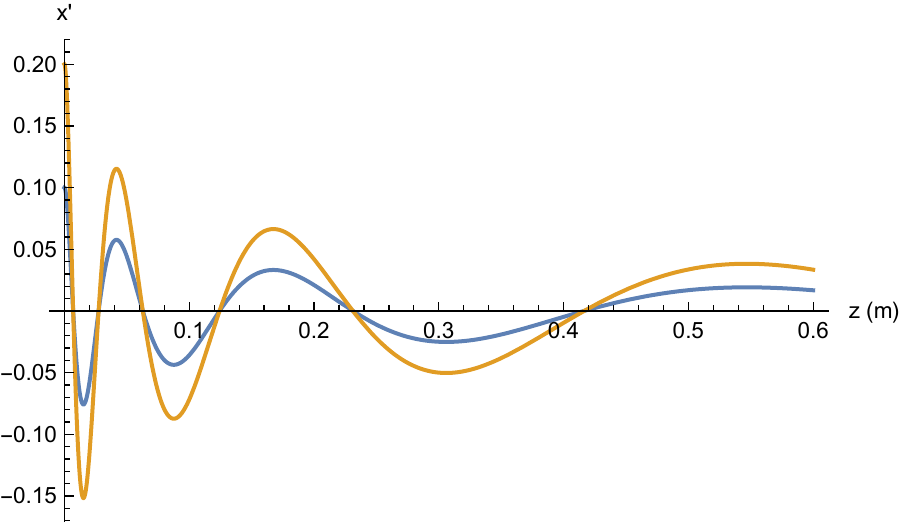} \\
\includegraphics[width=8.6cm]{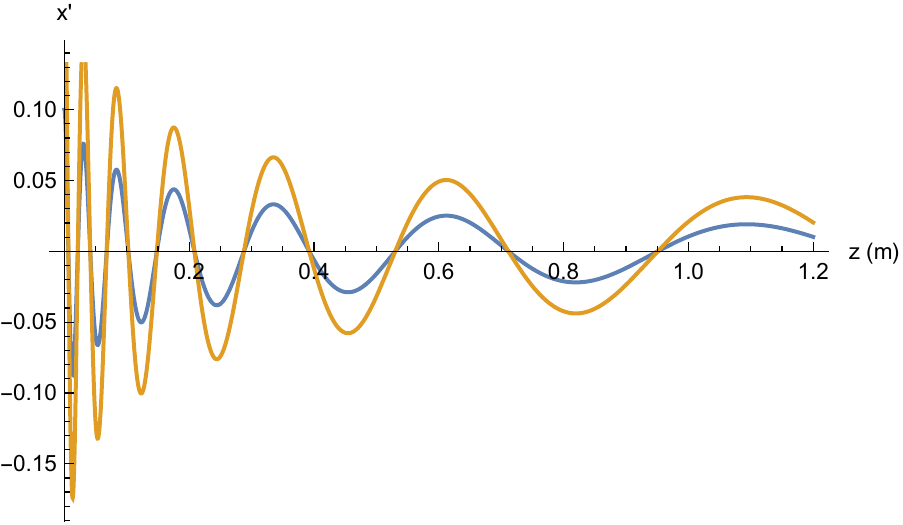}
\caption{Projection of the angle evolution, $\alpha = 48.3\,\text{m}^{-1}$ (top) and $\alpha = 24.15\,\text{m}^{-1}$ (bottom).  The blue curves for $\theta_\text{init} = 0.1$, the yellow ones for $\theta_\text{init} = 0.2$.  \label{fig:xprime48}}
\end{figure}

Corresponding analytical `trajectories' of the angle starting at $\theta_\text{init} = \sigma $ are presented in Fig.\ref{fig:xprime48} .

As it seems from both  the simulations and the analytics, the initial angular spread decreased by the rear end of AMD by the factor $1/\sqrt{1+\alpha L} $. This factor is approximately the same  for both the short and the long AMDs. 

\section{Results and Outlook}
The study on the Adiabatic Matching Device based on the adiabatic approach was done. The longitudinal and transversal probability density functions were derived under the assumption of the Rayleigh distribution of the initial angles, the point-like transversal density profile, and the Gaussian longitudinal density.  Our study shows that the device effectively reduces the angular spread in the positron bunch, which is the main goal of AMD . The results of study are as follows:
\begin{itemize}
\item The AMD provides axially symmetrical achromatic focusing of positrons (electrons). 
\item Reduction of the angular spread  is  proportional to square root of ratio of the initial field magnitude to its final value. It is independent of both the particle energy and the field magnitude.   
\item The transverse size  of the bunch at AMD exit is directly proportional to the angular reduction factor, the particles energy, and inversaly to the field magnitude.   
\item Longitudinal profile increase of the bunch is independent of both the energy of positrons and of the magnetic field magnitude. For the same reduction factor -- ratio of the initial to final field magnitude --  the bunch is prolonged proportionally to the AMD physical length. 
\end{itemize}
 
Recommendations for the AMD design are as following: The maximal field magnitude provides maximal efficiency of operation, that is, the maximal angular spread reduction at necessary radial dimension. 

It should be noted that the longer the AMD the better the adiabatic conditions and therefore the transverse focusing of the bunch. On the other hand, the longer the AMD the longer the the output bunch. Therefore, the AMD length should be at  acceptable minimum. The latter may be evaluated by means of simulations.      

\acknowledgments{
First of all, two of the authors (EB and VM) are deeply grateful to IJCLab provided us with shelter when we had fled Ukraine under war. They would like to acknowledge the IDEATE International Associated Laboratory (LIA) for its support. The authors are thankful to Dr. A. Ushakov for fruitful discussions. Work was done in the scope of INSPIRER project financed by the ANR (Agence Nationale de la Recherche) under Grant No: ANR-21-CE31-000. Vast part of analytics was done with the \emph{Wolfram Mathematica 12} package.
}

\providecommand{\noopsort}[1]{}\providecommand{\singleletter}[1]{#1}%

\end{document}